\documentclass[pra,superscriptaddress,showpacs]{revtex4}
\usepackage{amssymb}
\usepackage{amsfonts}
\usepackage{amsmath}

\newcommand{\ket}[1]{\ensuremath{| #1 \rangle}}

\begin{document}

\title{On determination of statistical properties of spectra from parametric level dynamics}
\author{Miros{\l}aw Hardej}
\affiliation{Center for Theoretical Physics, Polish Academy of Sciences, Aleja
Lotnik{\'o}w 32/44, 02-668 Warszawa, Poland}
\author{Marek Ku\'s}
\affiliation{Center for Theoretical Physics, Polish Academy of Sciences, Aleja
Lotnik{\'o}w 32/44, 02-668 Warszawa, Poland}
\author{Cezary Gonera}
\affiliation{Department of Theoretical Physics II, University of  \L\'od\'z,
ul.\ Pomorska 149/153, 90-236  \L\'od\'z, Poland}
\author{Piotr Kosi\'nski}
\affiliation{Department of Theoretical Physics II, University of  \L\'od\'z,
ul.\ Pomorska 149/153, 90-236  \L\'od\'z, Poland}

\begin{abstract}
We analyze an approach aiming at determining statistical properties of spectra
of time-periodic quantum chaotic system based on the parameter dynamics of
their quasienergies. In particular we show that application of the methods of
statistical physics, proposed previously in the literature, taking into account
appropriate integrals of motion of the parametric dynamics is fully justified,
even if the used integrals of motion do not determine the invariant manifold in
a unique way. The indetermination of the manifold is removed by applying
Dirac's theory of constrained Hamiltonian systems and imposing appropriate
primary, first-class constraints and a gauge transformation generated by them
in the standard way. The obtained results close the gap in the whole reasoning
aiming at understanding statistical properties of spectra in terms of
parametric dynamics.
\end{abstract}

\maketitle

\section{Introduction}
\label{sec:introduction}

One of the most characteristic features of quantum systems which exhibit
chaotic behaviour in the classical limit is an affinity of their spectral
properties to random matrices. The famous Bohigas-Giannoni-Schmidt conjecture
\cite{bohigas84} states that the statistics of distances between neighbouring
energy levels of a quantum system with chaotic classical limit is well
described by the one derived from the Random Matrix Theory (RMT)
\cite{mehta91}.

A vast numerical and experimental evidence \cite{haake00,stockmann99} in favour
of this hypothesis was collected during last twenty years. There are also
convincing theoretical arguments supporting it \cite{muller04,muller05}. In the
present paper we would like to reconsider one of the first theoretical
approaches initiated by Pechukas \cite{pechukas83} and further developed by
Yukawa \cite{yukawa85,yukawa86}. The original idea consisted in deriving
differential equations describing parametric level dynamics i.e.\ the evolution
of eigenvalues, when the parameter controlling the amount of chaos in the
system changes, and applying the rules of classical equilibrium statistical
mechanics to the flow described by the derived differential equations, treating
the parameter as a fictitious time in which the "evolution" takes place. As
observed by Yukawa the resulting dynamical system was Hamiltonian, hence
applying rules of equilibrium statistical mechanics was straightforward; the
equilibrium distribution should be given as the Boltzmann one,
\begin{equation}\label{boltz}
\rho=\mathcal{N}\exp(-\beta \mathcal{H}),
\end{equation}
where $\mathcal{H}$ is the Hamilton function of the system, $\beta$ - a
fictitious temperature (to be determined in some way from the initial data),
and $\mathcal{N}$ - an appropriate normalization constant. The dynamical
variables of the model, apart from the energy eigenvalues, involved also other
ones. Integration over them over the phase space led to the equilibrium
distribution of energy levels. As shown by Pechukas and Yukawa the resulting
distribution coincides with those provided by RMT for the ensemble of real
symmetric matrices with identically and independently distributed elements
(forming the so-called Gaussian Orthogonal Ensemble).

As appealing and straightforward as the above outlined approach might be, one
should not overlook some fundamental obstacles appearing when attempting to
formulate it in a more rigourous way. Let us summarize briefly the most
disturbing of them.

Both Pechukas and Yukawa started with the quantum Hamiltonian of the form
\begin{equation}\label{pert}
H=H_0+\lambda V,
\end{equation}
where, according to their original interpretation, a time-independent $N\times
N$ Hermitian matrix $H_0$ represented a quantum system enjoying integrable
classical limit, whereas $V$ was an integrability-breaking part making the
whole system classically fully chaotic when $\lambda$ attained appropriately
large values. Thus the ambitious program was designed to actually investigate
the transition between spectra of integrable ($\lambda=0$) and nonintegrable
($\lambda$ - large) cases. Thus, treating $\lambda$ as a fictitious time, one
faces a problem belonging to non-equilibrium rather than equilibrium
statistical mechanics, and usefulness of tools of the latter could be doubtful.
Moreover, as it is clear from (\ref{pert}), as $\lambda$ grows the motion is
unbounded, in particular the eigenvalues of $H$ grow indefinitely and, without
an additional scaling, no `equilibrium' distribution of eigenvalues of $H$ is
attained (although it might be that the statistics of distances measured in
units of the mean distance approaches some `equilibrium'). There is a way of
curing the situation - one should turn to dynamics of rescaled energy levels,
what in fact consist in changing the $\lambda$-dependence of $H$ (see
\cite{haake00}). In the following we take another route - we would like to
investigate the spectra of unitary evolution propagators instead of Hermitian
Hamiltonians. Propagators being unitary have their spectra confined to the unit
circle independently of their specific parameter dependence. The idea goes back
to Dyson \cite{dyson62}, although in slightly different context - he realized
that it is more convenient to define probabilistic measures on ensembles of
unitary rather than Hermitian matrices, what eventually led to the definitions
of the circular ensembles of RMT.

In our case we do not pretend to mimic the same way. Instead we propose to
start with the propagator for a particularly simple time-dependent quantum
system in which the integrability breaking part in (\ref{pert}) has a form of
periodic instantaneous kicks, so the whole Hamiltonian reads now:
\begin{equation}\label{del}
H(\lambda)=H_0+\lambda V\sum_{n=-\infty}^{\infty}\delta(t-nT),
\end{equation}
where $V$ is some constant, Hermitian matrix and $T$ the period of the
perturbation.

The unitary evolution operator of the system (the propagator) is given as the
solution of:
\begin{equation}\label{schr}
i\hbar\frac{dF}{dt}=H(\lambda)F, \quad F(0)=I.
\end{equation}
Since the time dependence is periodic the whole information about the evolution
of the system is encoded in $F$ evaluated at time $T$, i.e.\ the propagator
transporting the system in time over one period of the perturbation. Since we
will be concerned with properties of only this particular period-one propagator
we will use $F$ to denote it without risking confusion with the general
time-dependent one. In case it is needed we will write $F(\lambda)$ to remind
its dependence of the perturbation parameter $\lambda$ inherited from the
original Hamiltonian (\ref{del}). In the following we put $\hbar=1$ and $T=1$.

In the case of a kicked system, the one-period evolution operator $F$ takes a
particulary simple form
\begin{equation}\label{Ukick}
F(\lambda)=\exp\left(-i\lambda V\right)F_0, \quad F_0:=\exp(-iH_0).
\end{equation}

Of our interest will be the eigenphases $\varphi_n(\lambda)$ (i.e.\ the phases
of the eigenvalues), sometimes called also quasienergies of $F(\lambda)$,
\begin{equation}\label{eigU}
F(\lambda)\ket{\phi_n(\lambda)}=
\exp\left(i\varphi_n(\lambda)\right)\ket{\phi_n(\lambda)},
\end{equation}
where $\ket{\phi_n(\lambda)}$ are eigenvectors of $F(\lambda)$.

Symmetries of the system in question (in particular with respect to the
reversal  of time) enforce additional symmetry conditions on $H$ and $F$.
Without any other symmetries present, $H$ is a general hermitian and $F$ a
general unitary $N\times N$ matrix (we will be mostly concerned with this
case), whereas eg.\ for integer-spin time-reversal-symmetric systems $H$ is
real symmetric $H^*=H=H^T$ and we can assume $F=F^T$ (i.e.\ $F$ symmetric).

As in the autonomous case (\ref{pert}) considered by Pechukas and Yukawa one
can derive a closed set of differential equations, describing a Hamiltonian
motion in a multi-dimensional phase-space with $\lambda$ as a fictitious time
(\cite{haake00,ksh87,nakamura87}) (see next section). Since eigenvalues of
an unitary matrix lie on the unit circle, their phases are restricted to an
interval $[0,2\pi[$, and in contrast to the autonomous case (\ref{pert}), the
motion of the relevant variables is now bounded. This observation eases to some
extend the above mentioned problems with the Pechukas-Yukawa approach, although
does not warrant any equilibration as $\lambda$ grows. The problem of attaining
the equilibrium was thoroughly discussed in \cite{bghkz01} where it was shown
that the time of effective equilibration goes to zero when approaching classical
limit of the systems in question, hence the equilibrium statistical properties
of the spectra should be detectable for large quantum numbers. We will
recapitulate this discussion in the following sections.

When passing to the distribution of eigenphases from the equilibrium
distribution (\ref{boltz}) involving all dynamical variables one should choose
an appropriate measure for integrating out the irrelevant variables. The
natural measure is inherited from the symplectic structure underlying the
parametric dynamics of the eigenphases and was discussed in \cite{hzkh01}.

Finally let us introduce briefly the main objective of the present paper aiming
at closing the final gap in vindication of the parameter dynamics approach to
statistical properties of quasi-energies outlined above. The postulated
Boltzmann distribution can be validated only if there are no other constants of
the motion apart from the Hamilton function itself, or in other words, when the
motion is ergodic on the whole constant-energy surface. In the case of
Pechukas-Yukawa parametric dynamics, as well as in the case of parametric
motion of the kicked periodic systems (\ref{del}) it is not the case. The
dynamical systems governing the parametric motion of eigenvalues or eigenphases
are so called generalized Calogero-Moser or Sutherland-Moser systems
\cite{wojciechowski85}. They posses many additional integrals of motion and in
fact the motion is ergodic on a much smaller invariant manifold \cite{k88}. The
simplest way to include the influence of additional integrals of motion
consists of using in place of canonical ensemble measure (\ref{boltz}) its
grand-canonical generalization
\begin{equation}\label{gencan}
  \rho\propto \exp{\Big\{-\sum_\mu\beta_{\mu}
  I_{\mu}\Big\}}\,;
\end{equation}
nailing down the invariant manifold on which the motion takes place by fixing
constants of the motion $I_{\mu}$ in the ensemble mean with the help of
Lagrange parameters $\beta_{\mu}$; one of these $I_{\mu}$ should be the
Hamilton function $\cal H$. Using the microcanonical ensemble $\rho\propto
\prod_\mu \delta(I_\mu-\overline{I}_\mu)$ fixing the values of the constants of
motion $I_\mu$ to their initial values $\overline{I}_\mu$ exactly, rather in the
ensemble mean as (\ref{gencan}) does, would be even more appropriate, but
technically more complicated - see \cite{haake00}, Chapter~6 for a discussion
of the problem.

Integration of $\rho$ over all dynamical variables except the eigenphases
yields the desired distribution $P(\varphi_1,\ldots,\varphi_N)$. Such a program
was performed in \cite{dietz89,dietz94} where it was shown that inclusion of
additional known integrals of motion leads to corrections of the order $1/N$ in
comparison with the predictions of RMT. This result is highly satisfactory,
since one expects convergence to RMT in the limit when the dimension of the
matrix $N$ tends to infinity (which, in many models corresponds to classical
limit of the quantum system). The only remaining problem is whether the
integrals of motion taken into account in \cite{dietz89,dietz94} are all, which
are needed to fix (in the ensemble mean) the invariant manifold on which the
motion is ergodic, or in other words, what is the minimal set of independent
integrals of motion determining the invariant manifold (see also \cite{haake00}
for the formulation of the problem). The independence of the integrals used in
the above mentioned papers was investigated in \cite{mnich93}. In the present
paper we close the last gap by showing that they determine the invariant
manifold to the extent which is appropriate to infer the distribution of
eigenphases.

In Section~\ref{sec:parametric} we briefly derive the dynamical equations for
the parametric motion of eigenphases, whereas in Section~\ref{sec:statistical}
we discuss the statistical mechanics of the one-dimensional gas of eigenphases.
Section~\ref{sec:integrals} is devoted to additional integrals of motion: we
give their form appropriate for further applications and briefly discuss their
independence. The final Section~\ref{sec:poisson} offers the ultimate solution
of the problem in frames of the theory of constrained Hamiltonian systems.

\section{Parametric level dynamics}
\label{sec:parametric}

Let $W(\lambda)$ diagonalizes $F(\lambda)$,
\begin{eqnarray}\label{diag}
W(\lambda)F(\lambda)W^{-1}(\lambda)=e^{-i\Phi(\lambda)}={\rm
diag}(e^{-i\varphi _1(\lambda)},\ldots,e^{-i\varphi _N(\lambda)}), \\
\Phi(\lambda)={\rm diag}(\varphi _1(\lambda),\ldots,\varphi _N(\lambda)).
\end{eqnarray}
In the following we shall skip exhibiting the explicit $\lambda$-dependence
when possible. Since $F$ is unitary, so is $W$, $W^{-1}=W^\dagger$. In the case
of a general unitary matrix, the diagonalizing matrix $W$ is not unique even
after ordering the eigenphases: we can always left-multiply it by a diagonal
unitary matrix without altering the result (\ref{diag}).

Let's define following auxiliary matrices
\begin{eqnarray}
v&:=&WVW^{-1}=v^\dagger, \label{v} \\
l&:=&ie^{i\Phi}\left[v,e^{-i\Phi}\right]=-l^\dagger. \label{l}
\end{eqnarray}
>From (\ref{l}) and the diagonal character of $e^{-i\Phi}$ we have
\begin{equation}\label{lmn}
l_{nn}=0, \qquad v_{mn} = \frac{il_{mn}}{1-e^{i(\varphi_m - \varphi_n)}},\quad
n\neq m.
\end{equation}
Differentiating (\ref{diag}) over $\lambda$ we arrive at
\begin{eqnarray}
\frac{d\Phi}{d\lambda}&=&i(a-e^{-i\Phi}ae^{i\Phi})+v, \label{lax1} \\
\frac{dv}{d\lambda}&=&[a,v], \label{lax2} \\
\frac{dl}{d\lambda}&=&[a,l], \label{lax3}
\end{eqnarray}
where
\begin{equation}
a=\frac{dW}{d\lambda}W^{-1}.
\label{a}
\end{equation}
For reasons which soon will be clear in the following we shall use the notation:
\begin{equation}\label{canvar}
q_n:=\varphi_n, \quad p_n:=v_{nn},
\end{equation}
The diagonal part of the matrix equation (\ref{lax1}) reads
\begin{equation}\label{dqn}
\frac{dq_n}{d\lambda}=v_{nn}=p_n,
\end{equation}
whereas its off-diagonal part gives the off diagonal elements of $a$ in terms
of $v_{mn}$
\begin{equation}\label{amn}
a_{mn}=\frac{i{v_{mn}}}{1-e^{-i(\varphi_m - \varphi_n)}}, \quad m\ne n,
\end{equation}
what, upon (\ref{lmn}), gives:
\begin{equation}\label{avsl}
a_{mn}=-\frac{l_{mn}}{4\sin{}^2\frac{q}{2}}, \quad \quad m\ne n.
\end{equation}
By an appropriate choice of the diagonalizing matrix $W$ - see remark, below
(\ref{diag}), we can choose $a_{nn}=0$.

Eliminating with the help (\ref{lmn}) and (\ref{amn}) $v_{mn}$ and $a_{mn}$ in
favour of $l_{mn}$  we obtain from the diagonal part of (\ref{lax2})
\begin{equation}\label{dpn}
\frac{dp_n}{d\lambda}=-\sum_{k\ne n}l_{nk}l_{kn}{\cal V}^{\prime }(q_k-q_n),
\end{equation}
and, form (\ref{lax3}) for $n\ne m$,
\begin{equation}\label{dlmn}
\frac{dl_{mn}}{d\lambda }=-\sum_{k\ne m,n}l_{mk}l_{kn}\left( {\cal V}(q_n-q_k)-
{\cal V}(q_k-q_m)\right),
\end{equation}
where
\begin{equation}\label{pot}
{\cal V}(q)=-\frac{1}{4\sin{}^2\frac{q}{2}},
\end{equation}
and $^\prime$ in (\ref{dpn}) denotes the derivative with respect to the
argument.

Equations (\ref{dqn}), (\ref{dpn}), and (\ref{dlmn}) are Hamiltonian with
$\lambda$ treated as a fictitious time
\begin{equation}\label{hameqs}
\frac{dq_n}{d\lambda}=\left\{\mathcal{H},q_n\right\},\quad
\frac{dp_n}{d\lambda}=\left\{\mathcal{H},p_n\right\},\quad
\frac{dl_{mn}}{d\lambda}=\left\{\mathcal{H},l_{mn}\right\},
\end{equation}
and with the Hamilton function
\begin{equation}\label{ham}
{\cal H}=\frac{1}{2}\sum_{n=1}^N p_n^2+\frac{1}{2}
\sum_{n,m=1}^N l_{mn}l_{nm}{\cal V}(q_n-q_m)=\frac{1}{2}\mathrm{Tr}\,v^2,
\end{equation}
if we define the following Poisson brackets among
the dynamical variables $q_n,p_n$, and $l_{mn}$
\begin{equation}\label{pb1}
\left\{ p_m,q_n\right\} =\delta _{mn},\quad \left\{ p_m,p_n\right\} =\left\{
q_m,q_n\right\} =0,
\end{equation}
\begin{equation}\label{pb2}
\left\{ l_{mn},l_{ij}\right\} =\delta_{in}l_{mj}-\delta_{mj}l_{in}
\end{equation}
\begin{equation}\label{pb3}
\left\{ p_m,l_{kn}\right\} =\left\{ q_m,l_{kn}\right\}=0.
\end{equation}

It might be appropriate to mention that the system of dynamical equations for
the parametric motion of eigenvalues of a Hermitian matrix considered by Yukawa
\cite{yukawa85} has the same form as (\ref{dqn}), (\ref{dpn}), and (\ref{dlmn})
but with a different form of the potential, ${\cal V}(q)=-1/{q^2}$, the
`spatial' coordinates corresponding to eigenvalues, and slightly different
definitions of $v$ and $l$. For a unified treatment of parametric motion in
autonomous and kicked cases see \cite{hzkh01,khzh97}.

The system of equations (\ref{dqn}), (\ref{dpn}), and (\ref{dlmn}) can be
treated as describing dynamics (in the fictitious time $\lambda$) of a
one-dimensional gas of particles on the unit circle interacting mutually
\textit{via} the potential (\ref{pot}), but with evolving `coupling strengths'
$l_{mn}$ becoming thus additional dynamical variables.

\section{Statistical mechanics of the gas of eigenphases}
\label{sec:statistical}

As the phase-space trajectory of the fictitious gas evolves in time $\lambda$,
the original matrix $F(\lambda)$ changes within a one-parameter family. It is
that family rather than a single dynamical system (which has a fixed value of
$\lambda$) which exhibits random-matrix type spectral fluctuations. Indeed, if
the motion is ergodic (we shall show below, that it is indeed, although not on
the whole energy surface) implies that $\lambda$ averages of spectral
characteristics like the distribution of spacings between adjacent quasienergy
levels equal ensemble averages. Of importance is thus the minimal time interval
$\Delta \lambda$ needed for time and ensemble averages to become practically
equal. Obviously $\Delta\lambda \to \infty$ is sufficient, but not necessary
provision. During the evolution `particles of the gas' (i.e.\ in fact, the
eigenphases) undergo mutual collisions. Observe that since the potential in
(\ref{ham}) is repulsive, they usually do not cross (i.e.\ do not exchange
positions). Such a real crossing of two eigenphases would demand vanishing the
respective value of $l_{mn}$. Instead what is usually observed in the region of
parameter $\lambda$ corresponding to classically chaotic behaviour, are so
called avoided crossings when two neighbouring quasienergies approach a minimal
nonzero distance when $\lambda$ changes. In fact, as numerical experiments
show, for systems which are classically chaotic such avoided crossings are
abundant \cite{bghkz01}. Due to collisions the gas reaches state in which the
motion of particles consists of fluctuations in the vicinity of equilibrium. If
$H_0$ is integrable and chaos develops gradually after switching the
perturbation $V$ and increasing the coupling strength $\lambda$, the spectrum
of $F(\lambda)$ equilibrate to RMT predictions only after certain `relaxation
time' when the phase space regions of regular motion have shrunk to relatively
negligible weight. If, on the other hand, as shown in \cite{bghkz01}, $H_0$ and
$V$ are both non-integrable the initial state of the fictitious gas is already
close to equilibrium, and then the window $\Delta\lambda$ in question need not
be much larger than the collision time $\lambda_{\rm coll}$ of the gas, i.e.\
the mean distance of avoided crossings for a pair of neighbouring levels. It
was shown in \cite{bghkz01} that the time elapsing between consecutive
collisions scales as $N^{-\nu}$, $\nu>0$. A $\lambda$-average for the over a
window $\Delta\lambda\propto\lambda_{\rm coll}$ thus involves a family of
operators $F(\lambda)$ which all yield identical classical dynamics in the
limit $N^{-\nu}\to0$.

The most straightforward application of statistical mechanics is to employ the
canonical ensemble for the distribution of the dynamical variables $(q,p,l)$
\begin{equation}\label{can}
\rho(q,p,l)\propto \exp(-\beta{\cal H}(q,p,l)).
\end{equation}
A straightforward integration over Gauss-distributed $p$ and $l$ gives
precisely the eigenphase density of random-matrix theory \cite{mehta91}, i.e.
\begin{equation}\label{can-phi}
P(q_1,\ldots,q_N)=\int
d^Np\,d^{N(N-1)/2}l\;{\rm e}^{-\beta{\cal H}}\propto\prod_{m<n}|{\rm
e}^{-{\rm i}q_m}-{\rm e}^{-{\rm i}q_n}|
\end{equation}
As explained in the introduction the reasoning would be reasonable, if there
were no other integrals of motion beside $\cal H$ itself. In the case other
integrals of motion $I_\mu$ exist, the appropriate ensemble to use is the
generalized canonical ensemble (\ref{gencan}).

\section{Integrals of motion}
\label{sec:integrals}

The equations (\ref{dqn}), (\ref{dpn}), and (\ref{dlmn}) are clearly integrable
(they can be solved simply by diagonalizing $F$ at given $\lambda$ and
calculating appropriate matrix elements), so one should expect that there are
much more integrals of motion than the Hamilton function (\ref{ham}) itself.
Indeed from (\ref{lax1}) and (\ref{lax2}) we see that the quantities
\begin{equation}\label{im0}
I_{k_1m_1\ldots k_nm_n}=\mathrm{Tr}\left(v^{k_1}l^{m_1}\cdots
v^{k_n}l^{m_n}\right),
\end{equation}
are indeed constants of motion, i.e.
\begin{equation}\label{im1}
\frac d{d\lambda }I_{k_1m_1\ldots k_nm_n}=0,
\end{equation}
and should be taken into account when constructing the generalized canonical
ensemble (\ref{gencan}). It can be shown \cite{dietz89,dietz94,haake00}, that such
an ensemble yields the distribution of level spacings as well as low-order
correlation functions of the level density in common with random-matrix theory,
to within corrections of order $1/N$. The only problem is whether all integrals
nailing down an invariant manifold are of the form (\ref{im0}).

It is possible to show that only $N^2-N$ of such integrals are independent.
Indeed, let us briefly recall the reasoning presented in  \cite{mnich93} and
instead of $I_{k_1m_1\ldots k_nm_n}$ consider $N^2$ quantities
\begin{equation} \label{ckm}
C_{km}:=\mathrm{Tr}\left(e^{i\Phi}v^ke^{-i\Phi}v^m\right), \quad 0\leq k,m\leq
N-1.
\end{equation}
>From the definition of $l$ (\ref{l})
\begin{equation}
e^{i\Phi}ve^{-i\Phi}=v-il,
\end{equation}
i.e.
\begin{equation}
e^{i\Phi}v^ke^{-i\Phi}v^m=(v-il)^kv^m,
\end{equation}
hence $C_{km}$ are linear combinations of (\ref{im0}).
\begin{enumerate}
\item  Not all $C_{km}$ are independent. Indeed, trivially $C_{k0}=C_{0k}=\mathrm{Tr}V^k$.
\item There are no more independent integrals of this type (i.e.\ with $k\geq
N$ or $m\geq N$). Indeed from the Cayley-Hamilton theorem applied to the matrix
$v$, a $C_{km}$ with $k$ or $m$ larger than $N-1$ can be expressed a linear combination
of the basic ones (\ref{ckm}).
\end{enumerate}
Using (\ref{diag}), and (\ref{v}) we can rewrite $C_{km}$ as
\begin{equation}
\begin{array}{ll}
C_{km}
&=\mathrm{Tr}\left(WF^\dagger W^\dagger v^k WF W^\dagger v^m\right) \\
&=\mathrm{Tr}\left(F^\dagger (W^\dagger v^k W) F (W^\dagger v^m W)\right) \\
&=\mathrm{Tr}\left(F^\dagger V^k F V^m\right) \\
&=\mathrm{Tr}\left(e^{-i\lambda V}U_0V^k U_0^\dagger e^{i\lambda V} V^m\right) \\
&=\mathrm{Tr}\left(U_0 V^k U_0^\dagger V^m\right)
\end{array}
\end{equation}
In the basis in which $V$ is diagonal i.e.\ $V_{ij}=V_i\delta_{ij}$
this reduces to
\begin{equation}
C_{km}=\sum_{p,q=1}^N(V_q)^k(V_p)^m|(U_0)_{pq}|^2,\quad
\quad 0\leq k,m\leq N-1.
\label{ckm1}
\end{equation}
Now (\ref{ckm}) can be treated as a system of $N^2$ linear equations for $N^2$
unknown quantities $|(U_0)_{pq}|^2$, $1\leq p,q\leq N$, to be expressed in
terms of $N^2$ quantities $C_{km}$, $0\leq k,m\leq N-1$. For a generic $V$
i.e.\ when all its eigenvalues $V_i$ are different the system can be solved,
since the determinant of the coefficient matrix is a power of the Vandermonde
determinant $D_{V}$ constructed from $V_i$,
\begin{equation}\label{vderm}
D_{V}:=
\left|
\begin{array}{cccc}
   1    &  V_1   & \ldots  & V_1^{N-1} \\
   1    &  V_2   & \ldots  & V_2^{N-1} \\
 \vdots & \vdots & \ddots  &  \vdots   \\
   1    &  V_N   & \ldots  & V_N^{N-1} \\
\end{array}
\right|=\prod_{i<j}\left(V_i-V_j\right).
\end{equation}
Thus all $C_{km}$ can be expressed as linear combinations of $|(U_0)_{pq}|^2$.
The $N^2$ real numbers $u_{ij}:=|(U_0)_{pq}|$ fulfill $2N$ relations stemming from the
normalization of rows:
\begin{equation}
\sum_{j=1}^N u_{ij}^2=1, \quad i=1,\ldots,N,
\label{rows}
\end{equation}
and columns
\begin{equation}
\sum_{i=1}^N u_{ij}^2=1, \quad j=1,\ldots,N,
\label{cols}
\end{equation}
of the unitary matrix $U_0$. The number of independent relations is equal to
$2N-1$ and is less by one than the total number of equations in (\ref{rows})
and (\ref{cols}) since summing all equations in (\ref{rows}) over $i$ gives the
same as summing all equations in (\ref{cols}) over $j$, namely
$N=\mathrm{Tr}U_0U_0^\dagger$. Finally thus all $|(U_0)_{pq}|^2$ involve
$N^2-(2N-1)=(N-1)^2$ independent parameters and the number of independent
$C_{km}$ which can be used to determine them must be at least equal. Since
$C_{k0}$ and $C_{0k}$ do not depend at all on $U_0$, they can not be used to
determine $|(U_0)_{pq}|^2$. To do this we are left only with $C_{km}$, $1\le
k,m\le N$ which are exactly $(N-1)^2$ in number. The $U_0$ - independent
integrals $C_{k0}$, $0\le k\le N$, on the other hand, can be used to determine
$N$ independent parameters of $V$ (traces of its powers, or what is equivalent,
its eigenvalues), so they are also independent. Since, trivially, $C_{00}=N$,
we are left with the independent integrals of motion of the form:
\begin{equation}\label{independ1}
C_{k}:=C_{k0}=\mathrm{Tr}V^k, \quad k=1,2,\ldots,N-1,
\end{equation}
which are $N-1$ in number, and $(N-1)^2$ integrals
\begin{equation}\label{independ2}
C_{km}=\mathrm{Tr}\left(U_0V^kU_0^\dagger V^m\right), \quad k,m=1,2,\ldots,N-1,
\end{equation}
i.e.\ all together $N^2-N$ independent integrals of motion.

Let us now count how many variables we have in (\ref{dqn}), (\ref{dpn}), and
(\ref{dlmn}). The variables $q_n$ are real as eigenphases of the unitary matrix
$F$ and there are $N$ of them. Also $p_n$ as diagonal elements of a Hermitian
matrix $v$ are real, there are $N$ of them as well. Since $l$ is antihermitian
and off-diagonal there are $(N^2-N)/2$ matrix elements $l_{mn}$, but since they
are, in the case of a general unitary matrix, complex, we should count
separately their real and imaginary parts. Finally thus we have
$N+N+(N^2-N)=N^2+N$ real variables. Comparing this with the number of found
integrals of motion ($N^2-N$) we are tempted to think that, in a generic case,
invariant manifolds are of the dimension $(N^2+N)-(N^2-N)=2N$.

On the other hand, in the coordinate frame in which $V$ is diagonal,
$V_{ij}=V_i\delta_{ij}$, the motion described by $F(\lambda)=\exp(-iV)U_0$
involves only $N$ independent frequencies $V_k$, and takes place on an
$N$-dimensional torus (and is ergodic on it i a generic case when the
eigenvalues of $V$ are not rationally dependent). It seems thus, we are still
missing $N$ independent integrals of motion.

\section{Poisson structure and constraints}
\label{sec:poisson}

Before identifying missing integrals and determining their influence (or lack
of) on the distribution of eigenphases, let's consider more carefully the
proposed Hamiltonian formulation. First, observe that the definition of the
manifold on which the level dynamics takes place as parameterized by the
coordinates $q_n,p_n$ and $l_{mn}$ and equipped with the Poisson structure
(\ref{pb1})-(\ref{pb3}) is slightly flawed. From the definition (\ref{l}) of
$l$ we have $l_{nn}=0$, but this is inconsistent with the Jacobi identity which
must be fulfilled by (\ref{pb3}):
\begin{equation}\label{jac0}
\left\{l_{pq},\left\{l_{ik},l_{mn}\right\}\right\}+
\left\{l_{ik},\left\{l_{mn},l_{pq}\right\}\right\}+
\left\{l_{mn},\left\{l_{pq},l_{ik}\right\}\right\}=0.
\end{equation}
Indeed, substituting to the above $m=k$ and $n=i$ and using $l_{kk}=l_{ii}=0$
whenever they appear on intermediate steps, we arrive at an erroneous result
$\delta_{pq}l_{kq}+\delta_{iq}l_{pi}-\delta_{kq}l_{pk}-\delta_{pi}l_{iq}$.
instead of zero. Thus we are not allowed to put $l_{nn}=0$ from the very
beginning as equations defining our manifold. Instead, if we want to keep the
Poisson brackets (\ref{pb3}) we should change the definition (\ref{l}) to
\begin{equation}\label{lcorr}
l=ie^{i\Phi}\left[v,e^{-i\Phi}\right]+iL,
\end{equation}
where $L$ is an arbitrary, real, diagonal matrix, i.e. we introduced $N$
additional dynamical variables. To understand their meaning let us return to
the derivation of the dynamical equations by diagonalizing matrix $W$
(\ref{diag}), but this time we do not impose additional conditions on $W$, i.e.
we do not assume that the diagonal matrix elements $a_{nn}$ of
$a=dW/d\lambda\cdot W^{-1}$ vanish. Instead we allow them to be arbitrary
functions of $\lambda$. It should be clear (and indeed we will show that it is
the case), nothing really depends on the choice of $a_{nn}$, since nothing
concerning the eigenvalues should depend on the choice of the diagonalizing
matrix.

The resulting equations of motion are derived in the same way as previous ones
(\ref{dqn}), (\ref{dpn}), and (\ref{dlmn}). In fact only the third of them is
altered and reads now,
\begin{eqnarray}\label{dlmn-new}
\frac{dl_{mn}}{d\lambda }=&-&\sum_{k\ne m,n}l_{mk}l_{kn}\left( {\cal V}(q_n-q_k)-
{\cal V}(q_k-q_m)\right)\nonumber \\
&+&l_{mn}(a_{mm}-a_{nn})+l_{mn}(l_{nn}-l_{mm}){\cal
V}(q_m-q_n).
\end{eqnarray}

Equations (\ref{dqn}), (\ref{dpn}), (\ref{dlmn-new}) are again Hamiltonian with
the same Poisson structure (\ref{pb1})-(\ref{pb3}), but with a new Hamilton
function
\begin{equation}\label{ham-new}
{\cal H}=\frac{1}{2}\sum_{n=1}^N p_n^2+\frac{1}{2}
\sum_{n,m=1}^N l_{mn}l_{nm}{\cal V}(q_n-q_m)+\sum_j^Na_{jj}l_{jj},
\end{equation}
depending on $N$ arbitrary (in general `time-', i.e. $\lambda$-dependent)
functions $a_{nn}$. The quantities $C_{mn}$ (\ref{ckm}) are again integrals of
motion. In addition, we easily calculate that
\begin{equation}\label{Hl}
\left\{{\cal H},l_{nn}\right\}=0,
\end{equation}
so $l_{nn}$ are also constants of motion. In fact, as it is clear from the
previous considerations, nothing concerning the eigenphases depends on actual
values of $l_{nn}$. We can thus impose constraints, eg.
\begin{equation}\label{constr1}
l_{nn}=0, \quad n=1,\ldots,N.
\end{equation}

At this point it is instructive and in fact very natural to describe the
encountered situation from the point of view of Dirac's theory of constrained
Hamiltonian systems \cite{dirac64}. The conditions (\ref{constr1}) are so
called primary constraints (i.e.\ they are not obtained form the equations of
motion) and can be imposed only \emph{after} evaluating all Poisson brackets to
avoid the problems with the Jacobi identity mentioned at the beginning of the
present section. Further, one calculates easily that
\begin{equation}\label{invol}
\left\{l_{mm},l_{nn}\right\}=0.
\end{equation}
Together with (\ref{Hl}) it means that the consistency condition
\begin{equation}\label{consist}
\{\mathcal{H},l_{mm}\}+\sum_{n=1}^N a_{nn}\{l_{nn},l_{mm}\}=0,
\end{equation}
are identically fulfilled and no other constraints, neither primary nor
secondary, are produced, nor additional conditions are imposed on the functions
$a_{nn}(\lambda)$.

Due to (\ref{invol}) $l_{nn}$ are automatically first-class constraints (recall
that according to Dirac terminology a quantity is of first-class if its Poisson
brackets with all constraints vanish, see \cite{dirac64}, p. 18). The new
Hamilton function (\ref{ham-new}) involves as many arbitrary functions (in our
case these are functions $a_{nn}(\lambda)$), as there are independent primary
first-class constraints. On the other hand, first-class primary constraints
(\ref{constr1}) may be always used to produce a gauge transformation generated
by
\begin{equation}\label{gengauge}
G(\lambda)=\sum_{n=1}^N\theta_n(\lambda) l_{nn},
\end{equation}
i.e.
\begin{equation}\label{gauge-l}
l_{ij}\mapsto e^{i\theta_i(\lambda)}l_{ij}e^{-i\theta_j(\lambda)}.,
\end{equation}
with arbitrary
$\lambda$-dependent $\theta_{k}$, $k=1,\ldots,N$.

We expect that an initial physical state determined by initial values of the
phase-space variables $(q_n, p_n, l_{mn})$ determines also its all future
physical states. Since the Hamilton function (\ref{ham-new}) depends on $N$
arbitrary functions, the same may happen to the values of $(q_n, p_n, l_{mn})$
at latter times. But the only freedom is now given by the gauge transformation
(\ref{gauge-l}) connecting the variables describing the same physical state of
the system for different choices of the gauge. Hence particular physical
properties of the state (eg.\ statistical properties of the distribution of
positions, ie., in our case, eigenphases) should be gauge-independent, and in
fact they are, since the gauge transformation does not influence the relevant
variables $q_n$.

To be even more concrete in explaining the role of the gauge transformation for
the present problem let us observe that by assuming $l_{mm}=0$ we recovered the
previous count of the number of variables vs.\ dimension of the invariant
manifold, since the number of variables was first increased by $N$ by
introducing the diagonal elements of $l$ and then decreased by the same number
by imposing constraints equating them to zero. To fix the (still) remaining $N$
degrees of freedom let us observe that the gauge transformation (\ref{gauge-l})
does not change the integrals of motion (in particular the Hamilton function
itself) after reducing to the manifold determined by the constraints
(\ref{constr1}), retaining also the equations of motion in their original form.
The transformation is intimately related to the freedom of choice of the
diagonalizing matrix $W$ in terms of $a$, it leads to
\begin{equation}\label{gauge-a}
a\mapsto i\frac{d\theta}{d\lambda}+e^{i\theta}ae^{-i\theta}, \quad \theta:={\mathrm
diag}(\theta_1,\ldots,\theta_2).
\end{equation}
With the help of (\ref{gauge-l}) we can fix in an arbitrary way $N$ (more
precisely $N-1$, but one additional is determined by the choice of initial
point on the unit circle) phases of the variables $l_{nm}$. Let us summarize
\begin{itemize}
  \item number of variables: $N_{var}=N^2+2N$ (the old ones plus the (imaginary
  parts of) diagonal elements of $l$
  \item number of independent integrals $C_{mn}$: $N_{int}=N^2-N$
  \item number of constraints $l_{nn}=0$: $N_c=N$
  \item number of phases fixed by choosing a gauge $N_g=N$,
\end{itemize}
hence $N_{var}-(N_{int}+N_c+N_g)=N=$ dimension of the invariant manifold.

Now it is clear that integrals of motion $C_{mn}$ (\ref{ckm}) are the only
quantities which should be taken into account when determining the equilibrium
distribution. Indeed, as already mentioned the constraints and the gauge,
involving only $l_{mn}$, do not influence eigenphases, what is a direct
consequence of the independence of the eigenvalues on the choice of the
diagonalizing matrix. Moreover, our choice $l_{nn}=0$ reduces the Hamilton
function (\ref{ham-new}) to originally considered one (\ref{ham}) and the whole
reasoning which led, after integration out of $p$ and $l$ variables and
neglecting corrections of order $1/N$, to random matrix results for the
eigenphases, is fully vindicated.

\section{Acknowledgments}

We enjoyed fruitful discussions with Stefan Giller and Pawe{\l} Ma\'slanka.
The support by SFB/TR12 'Symmetries and Universality in Mesoscopic Systems'
program of the Deutsche Forschungsgemeischaft and Polish MNiSW grant no.\
1P03B04226 is gratefully acknowledged.

\end{document}